\documentclass[pra,twocolumn,floatfix,superscriptaddress,amsmath,citeautoscript,aps,longbibliography]{revtex4-1}
\pdfoutput=1
\usepackage[T1]{fontenc}\usepackage[latin1]{inputenc}
\usepackage{dcolumn,graphicx,color,booktabs,microtype,afterpage} \graphicspath{{Figures/}}
\usepackage[charter,greekuppercase=italicized]{mathdesign}
\usepackage[colorlinks,plainpages=false,linkcolor=blue,urlcolor=blue,citecolor=blue,pdfpagemode=UseNone,pdfstartview=FitBH]{hyperref}

\newcommand{\GD}{green dioptase}
\newcommand{\CSO}{Cu$_6$Si$_6$O$_{18}$}
\newcommand{\GDf}{Cu$_6[$Si$_6$O$_{18}]\cdot 6$H$_2$O}

\def\bS{{\bf S}}

\def\b0{{\bf 0}}

\def\vev#1{\langle{#1}\rangle}

\def\Tr{\mathrm{Tr}}
\def\non{\nonumber\\}

\begin{document}
\title{High-field spin-flop state in green dioptase}

\author{O.~Prokhnenko}
\thanks{Corresponding author: prokhnenko@helmholtz-berlin.de}
\affiliation{Helmholtz-Zentrum Berlin f\"{u}r Materialien und Energie, D-14109 Berlin, Germany}
\author{G. Marmorini}
\affiliation{Department of Physics and Mathematics,
Aoyama-Gakuin University, Sagamihara, Kanagawa 252-5258, Japan}
\author{S.~E.~Nikitin}
\thanks{Present address: Paul Scherrer Institute, Villigen PSI CH-5232, Switzerland}
\affiliation{Max Planck Institute for Chemical Physics of Solids, N\"{o}thnitzer Str. 40, D-01187 Dresden, Germany}
\affiliation{Institut f{\"u}r Festk{\"o}rper- und Materialphysik, Technische Universit{\"a}t Dresden, D-01069 Dresden, Germany}
\author{D. Yamamoto}
\affiliation{Department of Physics and Mathematics,
Aoyama-Gakuin University, Sagamihara, Kanagawa 252-5258, Japan}
\author{A.~Gazizulina}
\affiliation{Helmholtz-Zentrum Berlin f\"{u}r Materialien und Energie, D-14109 Berlin, Germany}
\author{M.~Bartkowiak}
\affiliation{Helmholtz-Zentrum Berlin f\"{u}r Materialien und Energie, D-14109 Berlin, Germany}
\author{A.~N.~Ponomaryov}
\thanks{Present Address: Institute of Radiation Physics,
Helmholtz-Zentrum
Dresden-Rossendorf, 01328 Dresden, Germany.}
\affiliation{Dresden High Magnetic Field Laboratory (HLD-EMFL), Helmholtz-Zentrum
Dresden-Rossendorf, 01328 Dresden, Germany} 
\author{S.~A.~Zvyagin}
\affiliation{Dresden High Magnetic Field Laboratory (HLD-EMFL), Helmholtz-Zentrum
Dresden-Rossendorf, 01328 Dresden, Germany}
\author{H.~Nojiri}
\affiliation{Institute for Materials Research, Tohoku University, Sendai, 980-8578, Japan}
\author{I.~F.~D\'{i}az-Ortega}
\affiliation{Institute for Materials Research, Tohoku University, Sendai, 980-8578, Japan}
\author{L.~M.~Anovitz}
\affiliation{Chemical Sciences Division, Oak Ridge National Laboratory, Oak Ridge, Tennessee 37831, USA}
\author{A.~I.~Kolesnikov}
\affiliation{Neutron Scattering Division, Oak Ridge National Laboratory, Oak Ridge, Tennessee 37831, USA}
\author{A.~Podlesnyak}
\affiliation{Neutron Scattering Division, Oak Ridge National Laboratory, Oak 
Ridge, Tennessee 37831, USA}
\date{\today}

\begin{abstract}
The high-field  magnetic properties and magnetic order of the gem mineral green dioptase \GDf~have been studied by means of single-crystal neutron diffraction in magnetic fields up to 21~T and magnetization measurements  up to 30~T. In zero field, the Cu$^{2+}$-moments in the antiferromagnetic chains are oriented along the $c$-axis with a small off-axis tilt. For a field applied parallel to the $c$-axis, the magnetization shows a spin-flop-like transition at $B^*=12.2$~T at 1.5~K. 
Neutron diffraction experiments show a smooth behavior in the intensities of the magnetic reflections without any change in the periodicity of the magnetic structure. 
Bulk and microscopic observations are well described by a model of ferromagnetically coupled antiferromagnetic $XXZ$ spin-$\frac{1}{2}$ chains, taking into account a change of the local easy-axis direction. 
We demonstrate that the magnetic structure evolves smoothly from a deformed N\'eel state at low fields to a deformed spin-flop state in a high field via a strong crossover around $B^*$. The results are generalized for different values of interchain coupling and spin anisotropy.
\end{abstract}
\maketitle

\section{Introduction}

Due to recent progress in solid-state chemistry, a lot of materials relevant to both practical and fundamental applications can be nowadays produced in the laboratory. 
However, there remain sample preparation conditions that are difficult or impossible to reach experimentally, especially for large samples, either because of extreme pressures and temperatures required or the time-scales needed. As a result, some compounds, especially in large single-crystal form, can be found only in nature. 
Natural minerals were the first materials whose magnetic properties were studied and used by mankind.  Recently, they have again drawn the attention of the magnetism community~\cite{Inosov}. 
Materials containing spin-$\frac{1}{2}$ ions or exhibiting a suppressed magnetic order have turned out to be a perfect playground for the study of quantum many-body effects such as dimensional crossover~\cite{Lake}, fractionalized magnetic excitations~\cite{Mourigal}, condensation of magnetic excitations~\cite{Giamarchi} and spin-liquids~\cite{Balents}. In this work we focus on high-field magnetic properties of green dioptase, a gemstone whose name and first description appeared more than two centuries ago~\cite{Huy}. 

\begin{figure}[tb!]
	\includegraphics[width=1.0\columnwidth]{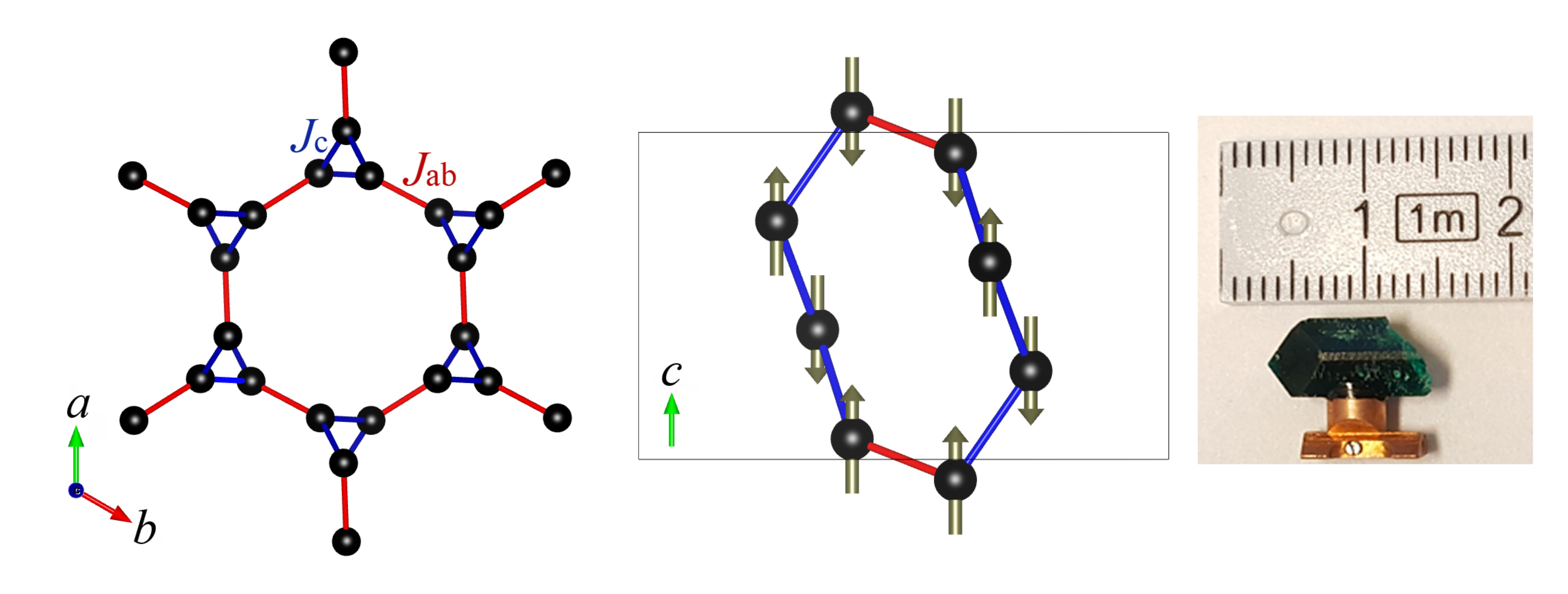}
	\caption {~(Left) Arrangement of Cu-ions in \GD~viewed along the $c$-axis. Both 
	intrachain (blue) and interchain (blue) exchange interactions are marked. (Middle) Schematic presentation of the magnetic structure viewed along the $a$-axis. For simplicity only two FM-coupled AFM-chains are shown within the crystallographic unit cell and the angle between the Cu-moments and $c$-axis is omitted. (Right) A photograph of the green dioptase single crystal fixed on a copper sample holder for the high-field neutron diffraction experiment.}
	\label{structure}
\end{figure}

The correct chemical composition of green dioptase, \GDf, was established in the 19th century~\cite{Hess}. 
Its rhombohedral crystal structure (SG $R\bar{3}$) is quite complex and consists of corrugated Si$_6$O$_{18}$ rings interconnected by Cu$^{2+}$ ions, see Fig.~\ref{structure}~\cite{Heide,Belokoneva,Ribbe,Breuer1989}. 
Analogous rings of H$_2$O molecules alternate with the silicate rings along the $c$-axis (hereafter we use the hexagonal notation). 
The magnetic spin-$\frac{1}{2}$ Cu$^{2+}$ ions, occupying a single Wyckoff site, form helical chains along the $c$-axis. 

The magnetic properties of green dioptase have been studied since the 1950s. 
Early NMR and specific heat measurement indicated that it orders antiferromagnetically (AFM) below 21~K \cite{Spence, Eisenberg}. 
This is close to the results from the most recent susceptibility, specific heat and neutron diffraction measurements reporting N\'{e}el temperatures of about 15~K~\cite{Ohta2009,Kiseleva,Belokoneva,Janson} and much lower than $T_\mathrm{N}$ obtained from susceptibility measurements~\cite{Newnham,Wintenberger}. 
The ordered magnetic moment stemming from the Cu$^{2+}$ ions is reduced and equals to $\sim 0.4-0.5 \mu_\mathrm{B}$~\cite{Wintenberger,Kiseleva,Belokoneva,Podlesnyak}. 

The arrangement of magnetic ions in green dioptase described above implies that each Cu-ion has two nearest neighbors along the chain and one in the $ab$-plane. 
Based on this geometry, two major exchange interactions are expected: intrachain, $J_c$, and interchain, $J_{ab}$ (Fig.~\ref{structure}). 
Neutron diffraction shows that the magnetic structure is AFM with propagation vector $\bf{k_M}=$(0,0,$\frac{3}{2}$). 
The magnetic moments point predominantly along the $c$-axis, but are inclined to it by about 13$^\circ$.
They are ordered AFM in the chains and ferromagnetically (FM) between them (Fig.~\ref{structure})~\cite{Belokoneva}. 
This is in agreement with the theoretical work of Janson~et~al.~\cite{Janson} and the inelastic neutron scattering report of Podlesnyak~et~al.~\cite{Podlesnyak}, which indicate $J_c > 0$ and $J_{ab} < 0$.
It is, however, in disagreement with the Quantum Monte Carlo calculations of Gros and co-workers who obtained only  AFM couplings~\cite{Gros}. 
Nonetheless, even if the right sign was obtained, the magnitudes of the derived exchange constants (as well as the ratio between them) vary significantly. 
For instance,  values of $J_c = $6.72 and 10.6 meV, and $J_{ab} = $-3.19 and -1.2 meV were reported in Refs.~\cite{Janson,Podlesnyak}, respectively.  
It is interesting to note that by annealing one can remove H$_2$O molecules from the crystal structure of \GD\ and produce \CSO\ phase. The dehydration significantly reduces the $J_{ab}/J_c$ ratio and brings the material to the one-dimensional regime, which is characterized by lower N\'{e}el temperature, smaller ordered moment of only 0.25~$\mu_{\mathrm{B}}$ and fractionalized spinon excitations~\cite{podlesnyak2019magnetic}.

Recently, Ohta et al. reported a spin-flop transition in green dioptase in a magnetic field of about 13~T applied along the $c$-axis at 1.5~K~\cite{Ohta2009}. 
Although a spin-flop transition would be expected in the case of the AFM chains in dioptase, the authors noted a difference in the slopes of the magnetization curves for $B\perp c$ and for $B\parallel c$ beyond 13~T. 
As single-ion anisotropy is not expected in the case of $S = \frac{1}{2}$ antiferromagnet, further investigations are required.  
In addition, the experimentally observed transition, which in the case of a classical spin-flop transition should be a sharp first order transition, is rather smooth. 
This is not the first time spin-flop transitions over a broad magnetic-field range have been observed, but this was previously attributed either to domain effects or to a misalignment of the applied magnetic field with respect to the AFM easy axis~\cite{Rohrer1975,King1979,Lynn1977}. 
On the other hand, intermediate phases between the AFM and spin-flopped states have been predicted theoretically~\cite{Yamashita1972,Liu1973,Becerra1974,Prystasz1982}.

All these issues question the nature of the transition at 13~T and the magnetic states around and above it in \GD. 
To-date there has been neither a direct confirmation of the spin-configuration at high fields nor information of their field evolution. 
The main experimental challenges here are the transition field is quite high, $B_c=12.5$~T, and the required field direction coincides with that of magnetic propagation $\bf{k_M}=$(0,0,$\frac{3}{2}$). 
This significantly restricts the number of available facilities in which such an experiment can be performed, especially as neutron scattering is needed to directly probe the microscopic alignment of the magnetic moments.  
In this paper we report a direct observation of the high-field phase in \GD\ using the High Field Facility for Neutron Scattering (HFM/EXED) at the BER II research reactor at Helmholtz-Zentrum Berlin (HZB)~\cite{ProkhnenkoJLSRF2017}. The unique combination of the High Field Magnet and the dedicated time-of-flight (TOF) Extreme Environment Diffractometer enables neutron scattering experiments in continuous magnetic fields up to 25.9 T and temperatures below 1~K~\cite{Smeibidl,ProkhnenkoRSI2015}. The neutron data are  supported by bulk magnetization measurements up to 30~T as well as Chain Mean Field Theory and Quantum Monte Carlo calculations. 
The experimentally revealed  spin-flop-like transition at $B^*=12.2$~T is interpreted in terms of the field-induced crossover from the low-field deformed N\'eel state to a high-field deformed spin-flop state, due to the involvement of several nonequivalent  magnetization sublattices. 

\section{Experimental details}

\begin{figure}[bt!]
\begin{center} 
\includegraphics[width=1.0\columnwidth]{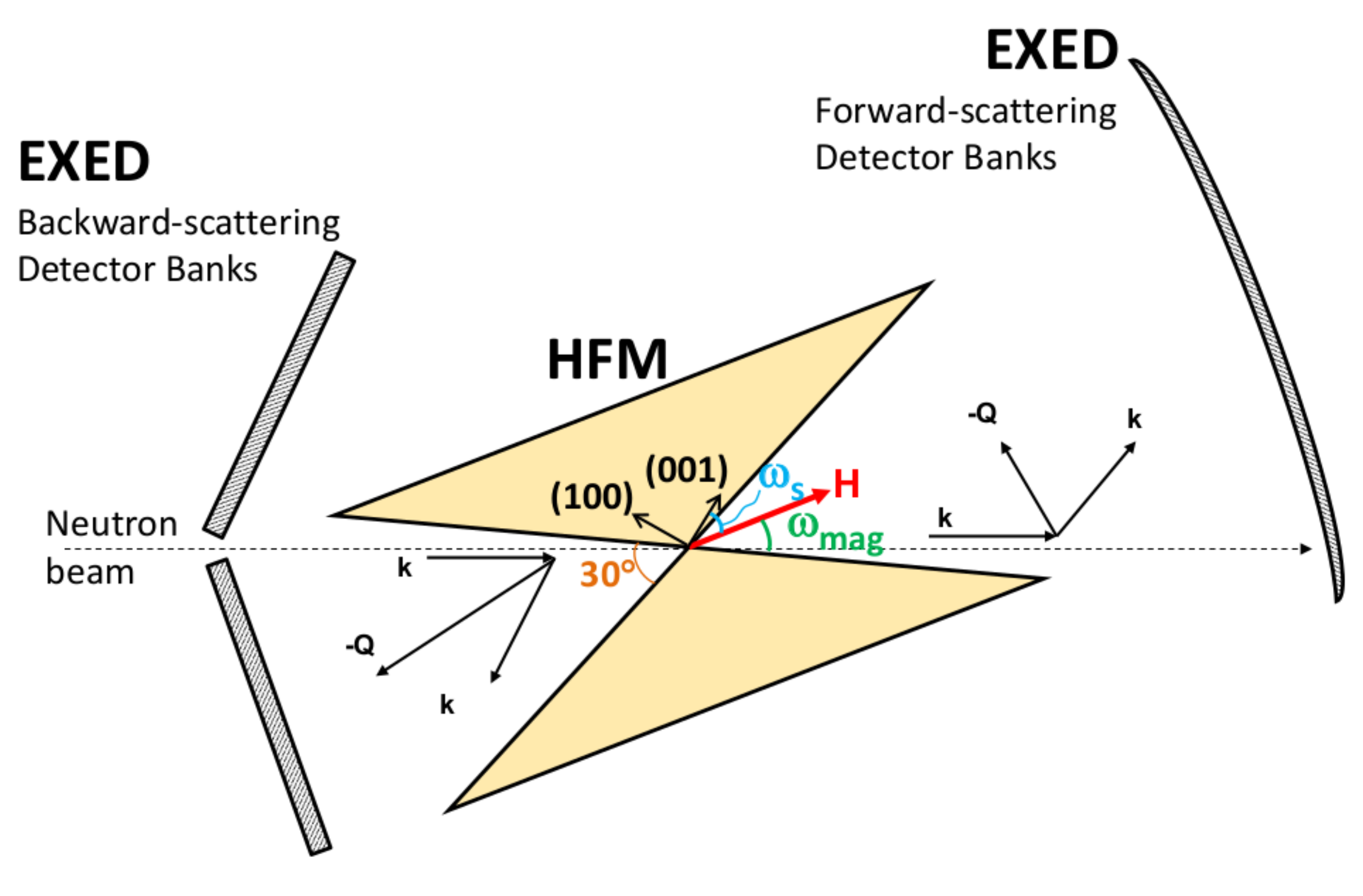} 
\caption{~Schematic presentation (top view) of the HFM/EXED setup. The picture includes the HFM and the EXED detectors, sample orientation and scattering geometry for the forward- and back-scattering detectors displayed for a 
single wavelength (wave vector $\bf k$). }
\label{scat_geom}
\end{center}
\end{figure}

The natural single crystals of \GD\ for the experiments described in this paper were obtained commercially. 
They originate from the Okawandasi Mine, Kunene Region, Namibia,  and Reneville, Brazzaville Department, Republic of Congo. 
These are the same crystals that were used in Ref.~\cite{Podlesnyak}. The samples were characterized by x-ray powder diffraction indicating a single phase. Absence of magnetic impurities was also confirmed by the magnetization measurements which show only the anomalies associated with the main phase. 

Bulk magnetic characterization of the samples was carried out by means of dc magnetic susceptibility and magnetization measurements up to 7~T using MPMS-3 at the Max Planck Institute in Dresden.  High-field magnetization measurements were performed using a 30~T pulsed magnet and a $^4$He flow cryostat at the Institute for Materials Research, Tohoku University (Sendai). For the magnetization measurements small crystals ($m=$17.6 and 14~mg) with dimensions of about 2~mm in length (along the $c-$axis) and about the same size in the basal plane have been used.

Single crystal neutron diffraction data in zero field were collected at the TOF Cold Neutron Chopper Spectrometer (CNCS)~\cite{CNCS1,CNCS2}, at the Spallation Neutron Source at Oak Ridge National Laboratory.
The crystal was aligned in the ($h$,0,$l$) scattering plane. 
The measurements were done at temperatures of $T=1.7$ and 25~K with a fixed neutron wavelength of $\lambda_{i}=4.96$~{\AA}. 

\begin{figure}[bt!]
	\begin{center} 
        \includegraphics[width=0.9\columnwidth]{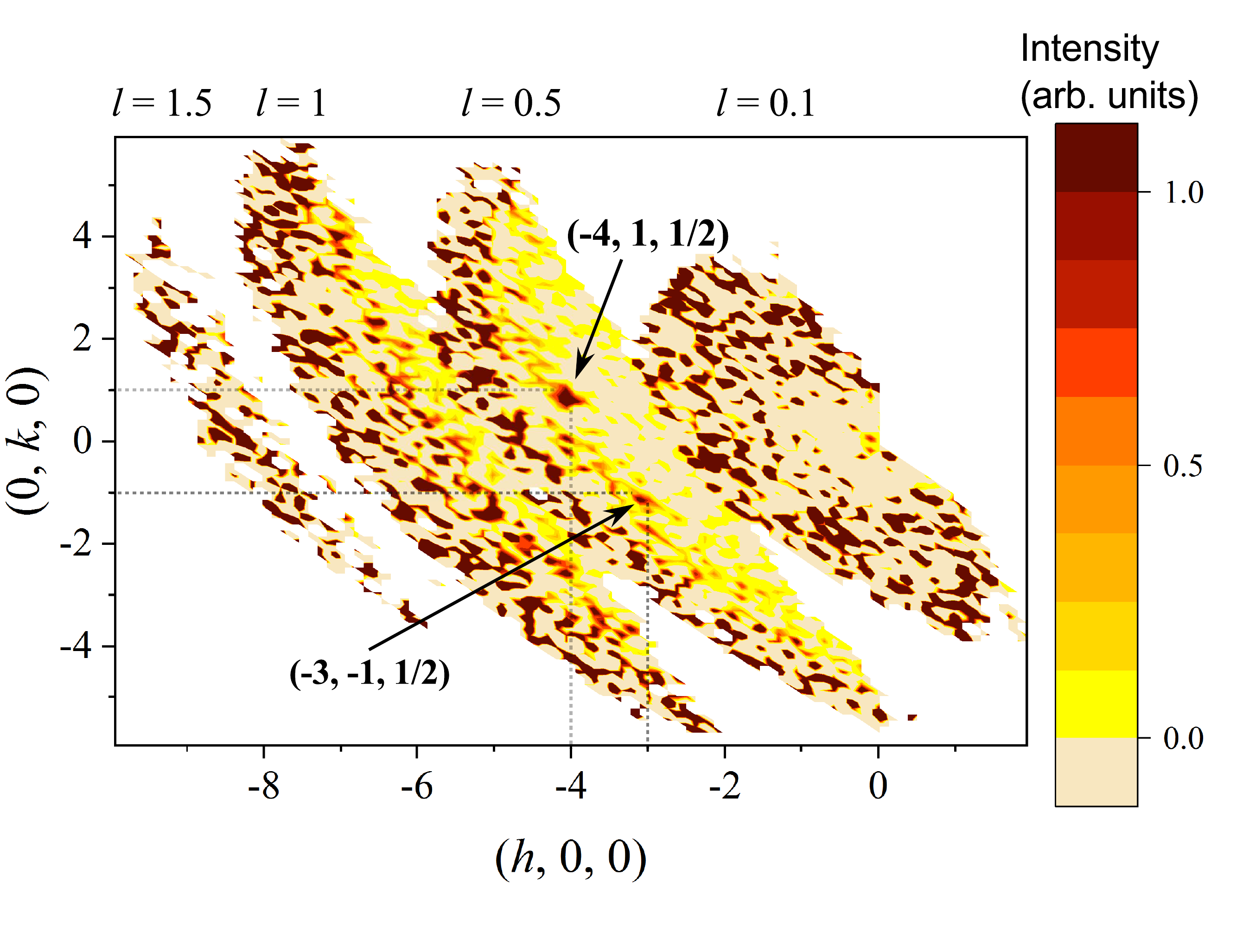}
		\caption{~Reciprocal ($h$,$k$)-maps for different $l$-values with 0.2$l$-width at zero field as covered in the HFM/EXED experiment. The data have been normalized to vanadium and the 21~T data have been subtracted to visualize the magnetic peaks.}
		\label{Qrange}
	\end{center}
\end{figure}

Neutron diffraction experiments in magnetic fields up to 21~T were carried out using the HFM/EXED high-field neutron facility at the BER-II research reactor at Helmholtz-Zentrum Berlin \cite{ProkhnenkoJLSRF2017}.  
The facility consists of a hybrid High Field Magnet (HFM) and a dedicated TOF Extreme Environment Diffractometer (EXED) \cite{Smeibidl,ProkhnenkoRSI2015}. The horizontal-field magnet has 30$^{\circ}$~conical openings on both ends, where the position-sensitive detectors are placed (Fig.~\ref{scat_geom}). 
In addition the HFM can be rotated with respect to the incident neutron beam by an angle of $\omega_{mag}\le 12^{\circ}$, which in combination with the TOF technique,  extends the reciprocal space coverage. 
For the current experiment the magnet was rotated by 11.85$^\circ$ deg with respect to the incident beam. 
The sample was mounted in a He-flow cryostat inserted into the room temperature bore of the magnet. 
The cryostat is equipped with a rotation stage around the vertical axis with an angular range of $\omega_{s}\approx 180^{\circ}$, allowing the sample orientation to be adjusted in-situ. 
The crystal was oriented such that the scattering plane was spanned by the vectors (1,0,0) and (0,0,1). The  $c$-axis was deliberately misaligned relative to the magnetic field, in order to access the magnetic reflections having finite component along the [0,0,$l$] direction. 
A rotation of $\omega_{s}=13^\circ$ around the vertical axis was applied for this purpose. A sketch of the scattering geometry is shown in Fig.~\ref{scat_geom}. The covered momentum transfer ($Q$) range presented as ($h$,$k$)-maps for selected $l$-values is displayed in Fig.~\ref{Qrange}. The data collection was performed with a fixed $\omega_{s}$ (i.e. fixed direction of the magnetic field with respect to the sample). All the measurements were performed at $T=1.4$~K.

For the neutron diffraction experiments much larger sample ($m=$0.6 g) with 11 mm along the $c-$axis and 5 mm across the diameter has been chosen (Fig.~\ref{structure}). Neutrons interact with matter weakly forcing to use samples with larger volume. Moreover, the ordered Cu-moment, the elastic neutrons scattering is sensitive to, is quite small for a $S = \frac{1}{2}$ system. 
To deal with the above issues the EXED instrument configuration was optimized to maximize the neutron flux on the sample for the given $Q-$range of interest. This is achieved by trading the wavelength bandwidth for the repetition rate and increasing the measurement time. 
For the current measurements the bandwidth was set to 0.7-2.65~{\AA} leading to the instrument repetition rate of 30~Hz. 
At each field the data collection constituted 3~hrs. The magnet ramping time at fields below 18~T was about 
0.5~T/min and 0.3~T/min above it.

The \textsc{Mantid} \cite{Mantid}, \textsc{FullProf} \cite{FullProf}, \textsc{BasIrreps} \cite{FullProf} and \textsc{Vesta} \cite{Momma} software packages were used for data reduction, analysis and visualization.

\begin{figure}[tb]
    \includegraphics[width=0.9\columnwidth]{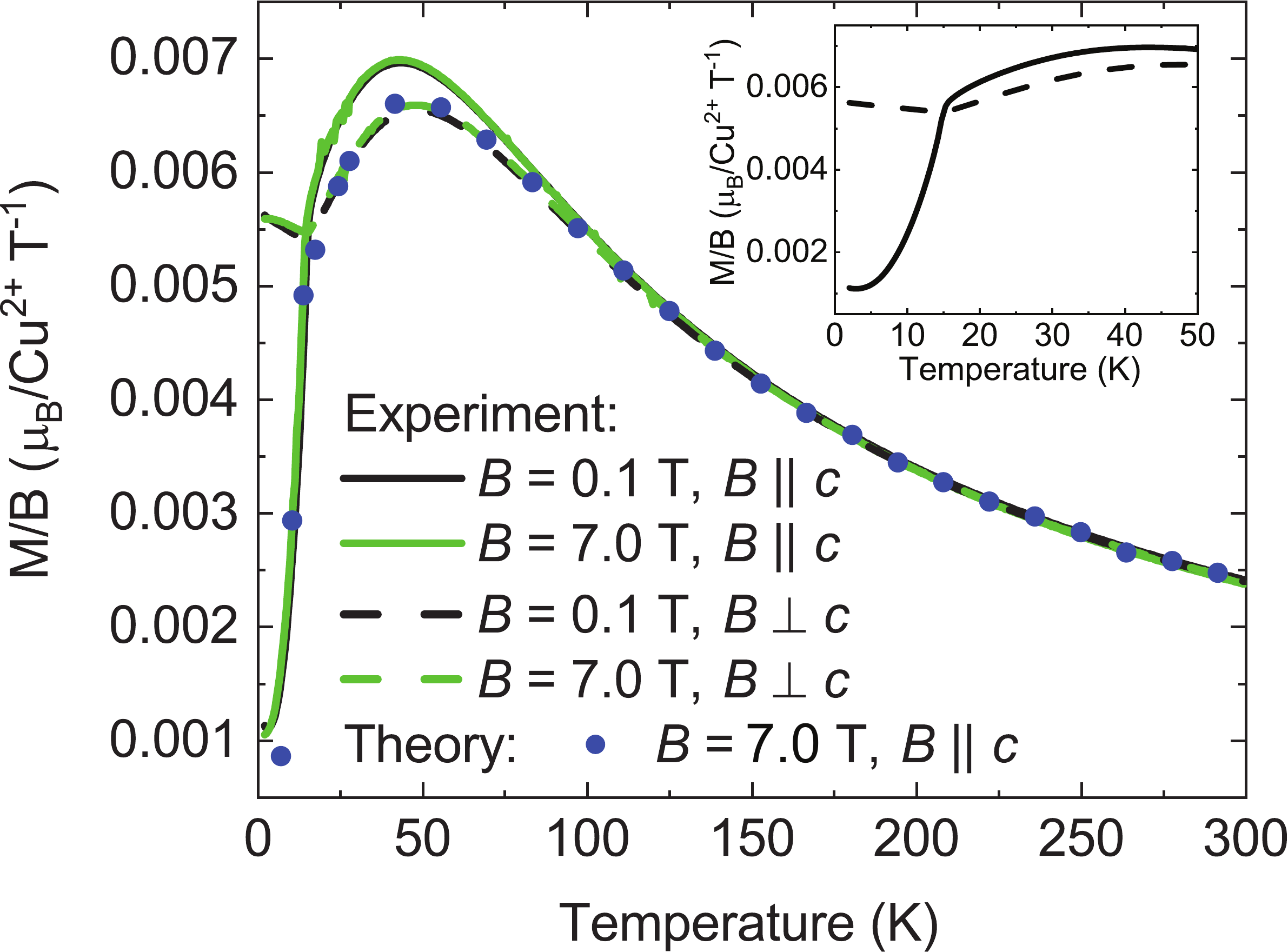}
    \caption{~Temperature dependencies of the static spin susceptibility of \GD~measured in magnetic fields of 0.1 and 7~T applied parallel and perpendicular to the $c$ axis. The theoretical calculation was done using a QMC simulation of a simplification of model Eq.~\ref{ourmodel} as explained in the text. The inset magnifies the low temperature part of the plot.}
    \label{MvsT}
\end{figure}

\section{Results} 
\subsection{Bulk Properties}

The magnetic properties of \GD\ have been reported by a number of authors~\cite{Newnham, Wintenberger, Gros, Ohta2009, Janson}. 
We have performed bulk characterization of our sample, and the results 
agree with those reported in Refs.~\cite{Gros, Janson}. Figure~\ref{MvsT} displays the temperature dependence of  the static spin susceptibility $M/B$ measured in magnetic fields of 0.1 and 7~T applied both parallel and perpendicular to the $c$-axis. Above approximately 150~K the data can be fitted with the Curie-Weiss law. The obtained fit parameters, the Weiss constant 48~K and the effective magnetic moment $\mu_{\mathrm{eff}} = $1.94 $\mu_{\mathrm{B}}$ ($g = $2.2), agree well with those reported in Ref.~\cite{Janson}. 
Around 100~K the susceptibility curves measured with the field applied along and perpendicular to the $c$-axis start deviating from each other, though they still show similar temperature dependencies (including a broad maximum around 45~K) down to $T_\mathrm{N}=$15~K. Below this temperature, the curve for $B\perp c$ shows an upturn, while  the susceptibility for $B\parallel c$ changes its slope and drops rapidly as detailed in the inset of Fig.~\ref{MvsT}. These data are in good agreement with the single crystal measurements reported by Gros~et~al.~\cite{Gros}.
 
\begin{figure}[tb]
	\includegraphics[width=0.9\columnwidth]{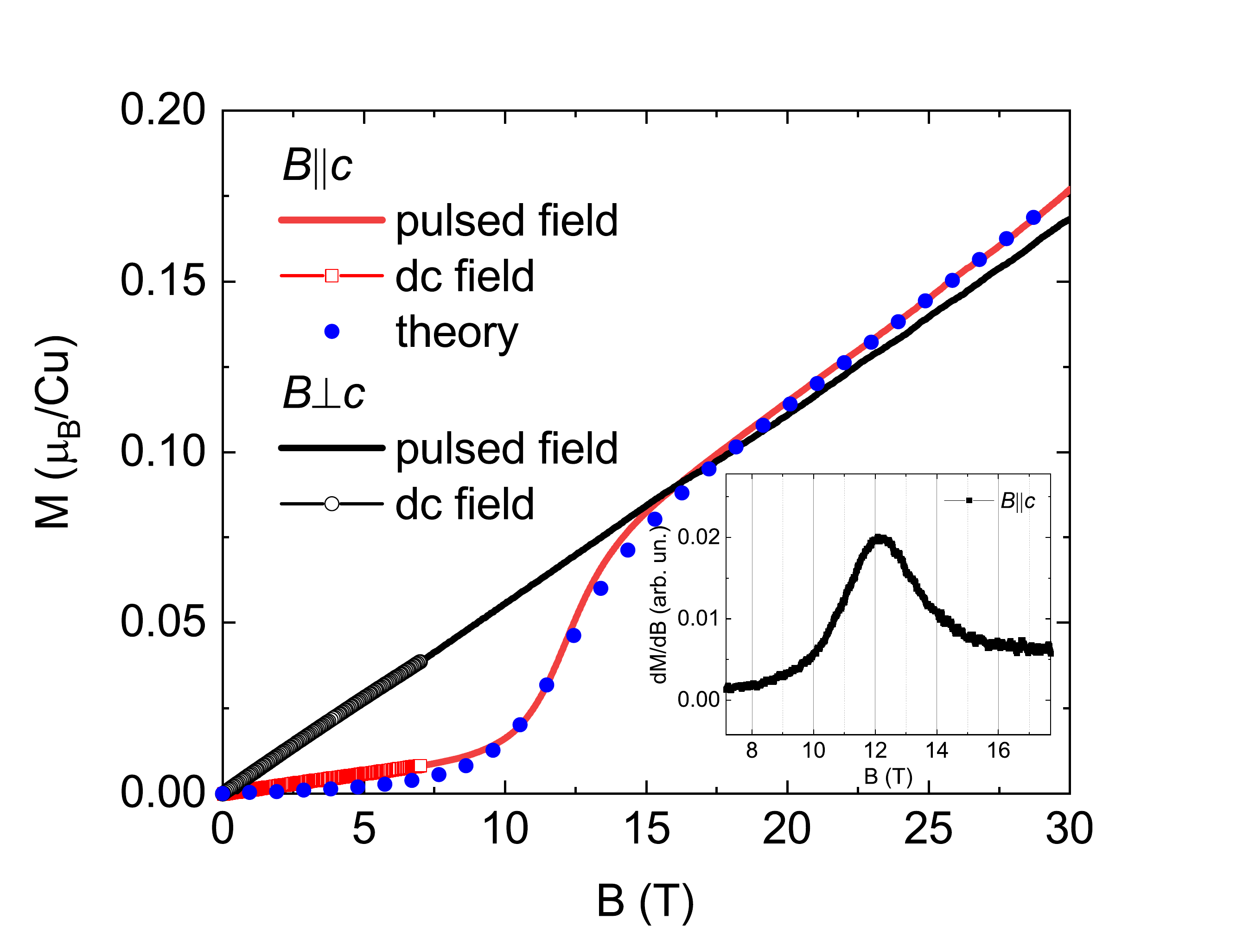}
    \caption{~Magnetization curves (solid line) of green dioptase measured at 1.5~K for a field applied parallel and perpendicular to the $c$-axis. The pulsed-field data have been normalized to the DC-field measurements at low fields. The theoretical magnetization curve was obtained from ChMFT simulations of model Eq.~\ref{ourmodel} at $T = 0$. The inset displays a first derivative of the experimental magnetization as function of field.}
    \label{MvsH}
\end{figure}

Figure~\ref{MvsH} displays the field dependence of the magnetization measured in pulsed magnetic fields up to 30~T applied parallel and perpendicular to the $c$-axis. The absolute value of the magnetization was checked against the low-field measurements on a SQUID magnetometer. 
For $B \| c$ there is a metamagnetic-like transition at about $B^*=12.2$~T.
This is slightly less than the transition field of 13~T reported by Ohta et al.~\cite{Ohta2009}. The inset in Fig.~\ref{MvsH} shows the first derivative of the magnetization, in which the transition field is clearly visible.

For $B\perp c$ the magnetization grows linearly with the field and does not show any saturation up to the highest field applied. Contrary to Ohta et al., however, the slope of the magnetization beyond 13~T for $B\perp c$ coincides with the one for $B\parallel c$ within the experimental precision of the sample alignment of 1-2$^\circ$.

\subsection{Neutron Scattering}

\begin{figure}[tb]
	\includegraphics[width=1.0\columnwidth]{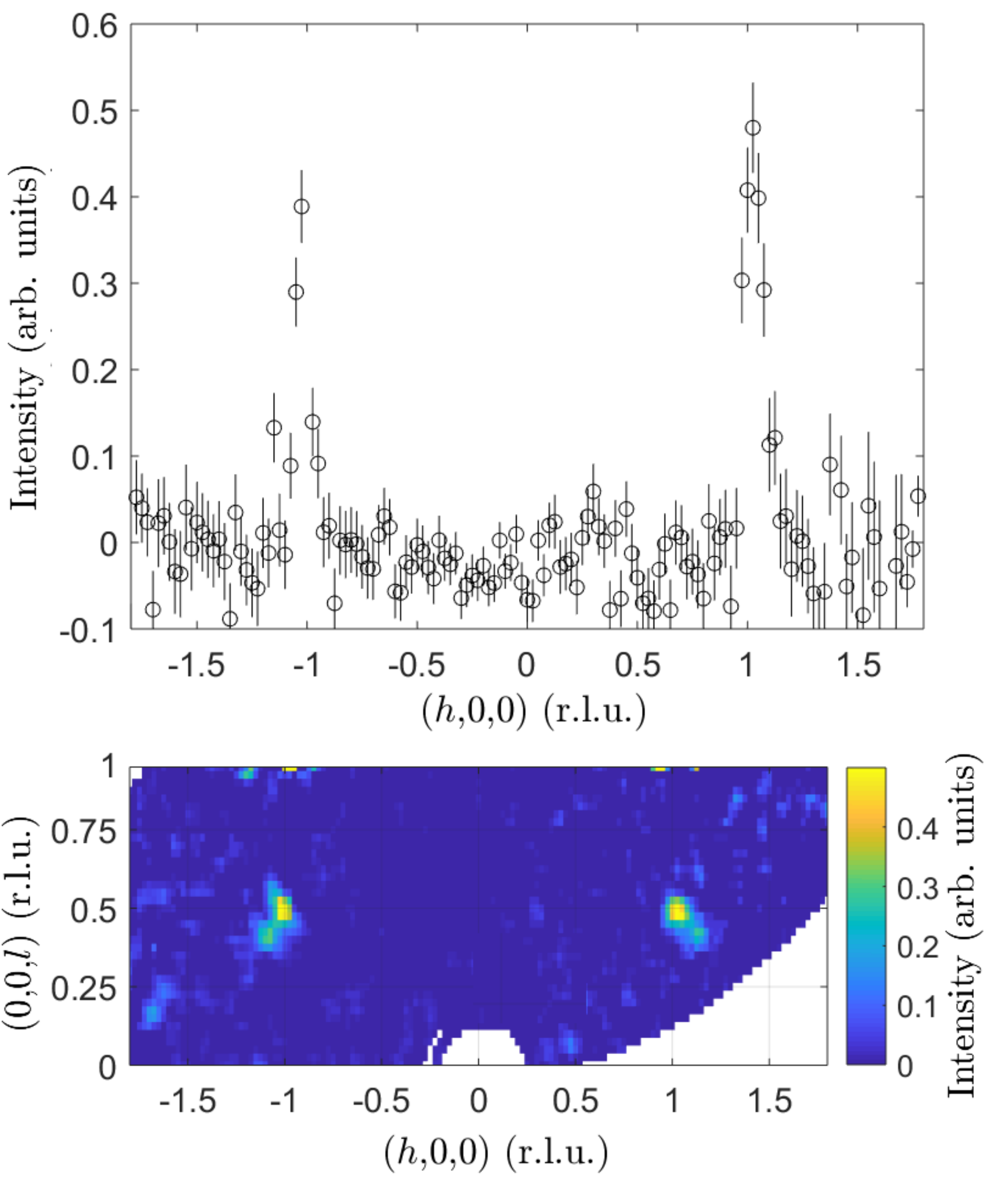}
    \caption{~The difference of elastic scattering intensity (integrated over energy $E$=[-0.1; 0.1] meV), obtained by subtracting the CNCS data sets measured at temperatures $T$=1.7 and 25~K. (top) The Q-cut along wave vector ($h$,0,$\frac{1}{2}$), integrated over $l$=[0.4; 0.6] r.l.u. and $k$=[-0.1; 0.1] r.l.u. (bottom) Contour plot of the magnetic scattering in the ($h$,0,$l$) plane, integrated over wave vector $k$=[-0.1; 0.1] r.l.u.}
    \label{slicecut}
\end{figure}

To get a microscopic insight into nature of the above metamagnetic transition we performed single crystal neutron diffraction in high magnetic fields. 
First, we checked the zero-field ground state of our sample using the CNCS instrument at ORNL. 
The pattern, taken at $T=1.7$~K, shows weak extra peaks not seen in the data measured above the N{\'e}el temperature at $T=25.0$~K, as illustrated in Fig.~\ref{slicecut}.
The observed magnetic reflections can be indexed with a magnetic propagation vector $\bf k=$(0,0,$\frac{3}{2}$) in agreement with the AFM structure reported by Belokoneva et al. \cite{Belokoneva}.

\begin{figure*}[tb]
    \includegraphics[scale=0.5]{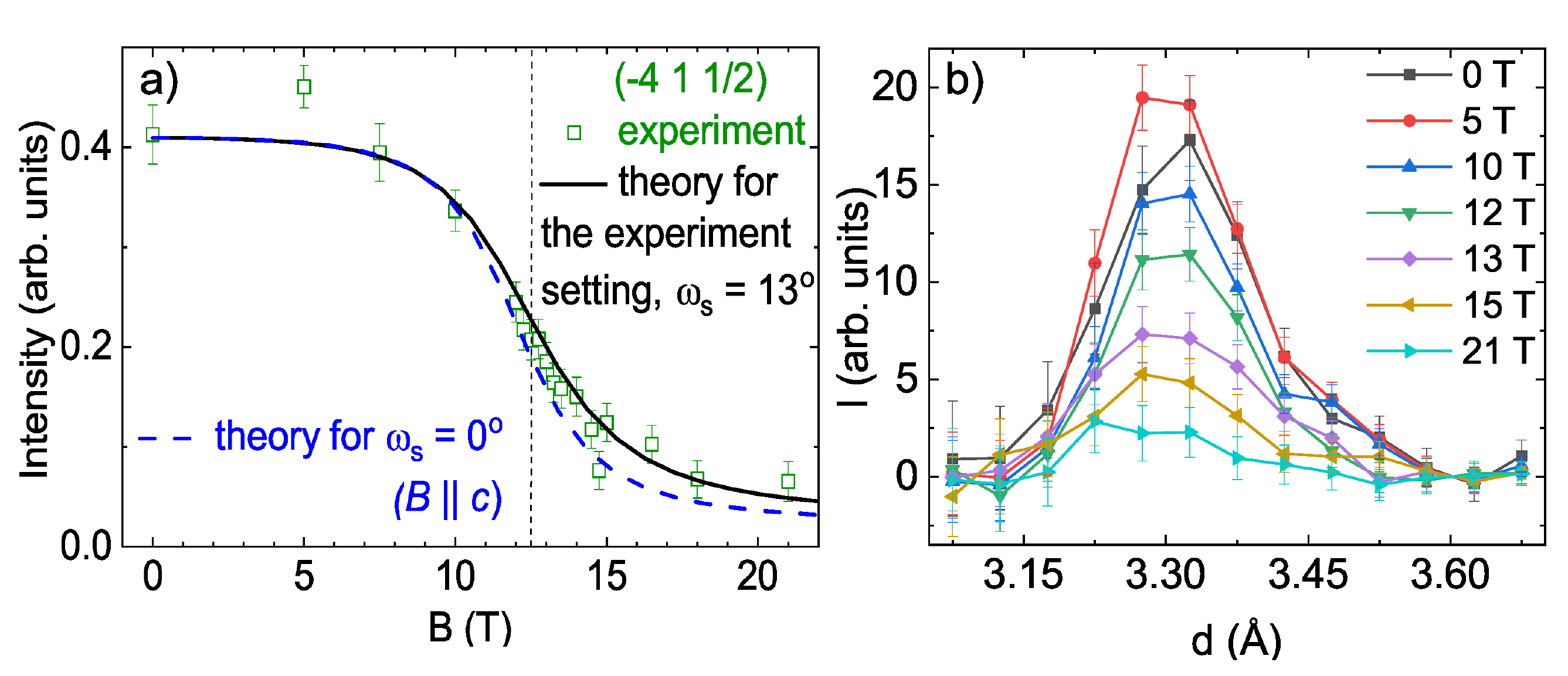}
    \caption{~(a) Intensity of the (-4, 1, $\frac{1}{2}$) magnetic reflection as function of magnetic field. The plot shows both the experimental data (squares) and the theoretical calculations for both the experimental setting, $\omega_s = 13^\circ$ (line), and the setting $B\parallel c$, $\omega_s = 0^\circ$ (dashed line). The vertical dash line corresponds to the transition field corrected for the experimental angular offset  $\omega_{s}$=13$^\circ$. (b) A profile of the (-4, 1, $\frac{1}{2}$) reflection at  different fields.}
    \label{mag_ref}
\end{figure*}

For the high-field measurements on HFM/EXED the crystal was aligned such that the horizontal field made an angle of $\omega_s=13^\circ$ with the $c$-axis ([0,0,1]-direction). Details of the experimental setup are presented in Fig.~\ref{scat_geom}. In this configuration we observed two reflections appearing below $T_\mathrm{N}$, which can be indexed as (-4,1,$\frac{1}{2}$) and (-3,-1,$\frac{1}{2}$) as well as nuclear reflections (-7,3,1), (-6,1,1) and (-5,-1,1). 
As most of  the nuclear reflections appear on the edge of our $Q$-coverage (see Fig.~\ref{Qrange}), they could be used only for sample orientation purposes. 
Figure~\ref{mag_ref}a shows the field dependence of the intensity of the stronger (-4,1,$\frac{1}{2}$) magnetic reflections. 
The plot shows that beyond 8~T  the intensity of this reflection decreases rapidly through the transition. Above 16~T the intensity of (-4,1,$\frac{1}{2}$) reflection levels off and remains finite. 
The intensity of the (-3,-1,$\frac{1}{2}$) reflection also decreases with the field. However, it is about five times weaker the (-4,1,$\frac{1}{2}$) reflection, resulting in noisier data preventing us from obtaining its accurate field dependence. 

As previously noted with respect to the magnetization data, the transition does not look abrupt. 
Instead, the variation of intensity occurs over a wide magnetic field range. 
In addition, the intensity dependence is not accompanied by changes in the positions of the reflections (Fig.~\ref{mag_ref}b) and no extra reflections have been observed above the transition in the $Q$-range covered (see Fig.~\ref{Qrange}).
 
\begin{figure*}[t!]
	\begin{center}
	\includegraphics[width=1.9\columnwidth]{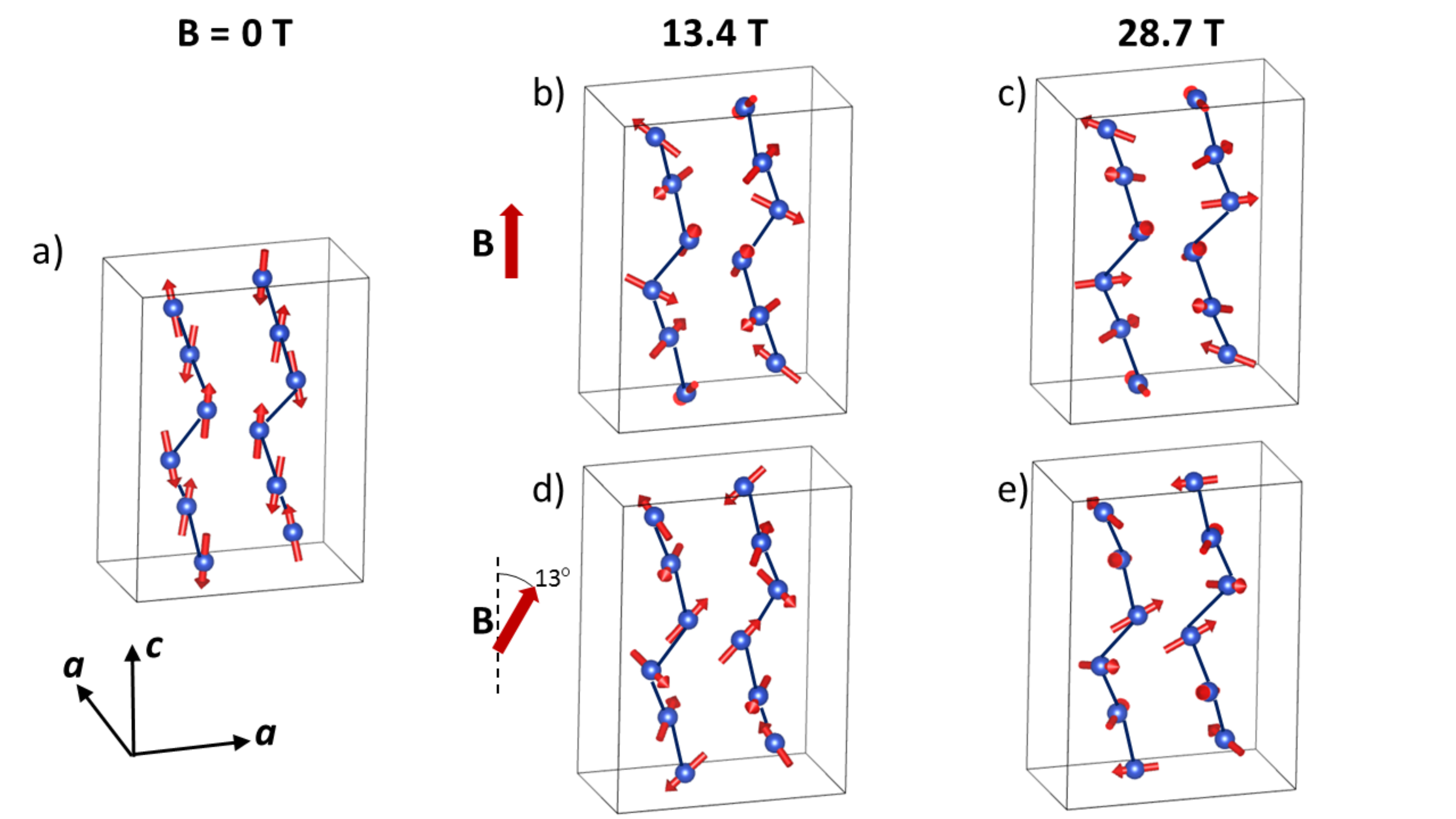}
	\caption {~Magnetic unit cell of \GD~at zero field (a), across the transition, $B=13.4$ T (d), and above the transition, $B=28.7$ T (e) for the experimental setting $\omega_s = 13^\circ$. For comparison, the panels (b,c) show  high-field magnetic structures for a setting $B\parallel c$. For better visibility, only two Cu-chains are shown. }
	\label{structure2}
	\end{center}
\end{figure*}

According to Belokoneva~et~al.~\cite{Belokoneva}, the ground state AFM structure consists of FM-coupled chains in which the Cu-moments are aligned predominantly along the $c$-axis (see Fig.~\ref{structure}). Table~\ref{Irreps} contains the irreducible representations (irreps) which can be used to define possible magnetic structures with  $\bf k$=(0,0,$\frac{3}{2}$). 
Those that allow FM coupling between the chains are $\Gamma_1$, $\Gamma_3$ and $\Gamma_5$. Among them the irreps $\Gamma_3$ and $\Gamma_5$ lead to amplitude modulated structures and thus can be neglected. 
As a result, the experimentally determined magnetic structure~\cite{Belokoneva} is in agreement with irrep $\Gamma_1$ with a dominant S$_\parallel$ ($\parallel c$) spin-component. Important to note here that the moments are tilted off the $c$-axis by 13$^\circ$ making finite in-plane S$_{\perp}$ spin-components. 
Because of the crystal symmetry, the in-plane spin component rotates by 120$^\circ$ when going from one layer to the adjacent one along in the chains while keeping the handedness of the helicity in neighboring chains opposite. 
The magnetic unit cell is shown in Fig.~\ref{structure2}a on the left hand side. 

As the propagation vector does not change above the transition, the same irreps can, in principle, be used to determine the high-field structure. Indeed, a change of the interchain coupling from FM to AFM is only physically possible if there is a structural change, which we do not expect to occur in green dioptase when a magnetic field is applied. 
However, as the field-induced structure must possess a ferromagnetic component in field direction, one has to consider a combination of the above irreps with those for $\bf{G_k}$ with $\bf k$=(0,0,0). In our case the little group of the propagation vector $(0,0,\frac{3}{2})$ is the full group $R\bar{3}$, meaning that the same Table~\ref{Irreps} holds for both cases. We come back to this issue in the next section.

\begin{table}
\caption{ \label{Irreps} Irreducible representation of the little group $\bf{G_k}$ containing all the symmetry elements which leave $\bf k$=(0,0,$\frac{3}{2}$) (and $\bf k$=(0,0,0)) invariant.}
\begin{ruledtabular}
    \begin{tabular}{ccccccc}
                    & 1 & $3^+$ & $3^-$ & $\bar{1}$ & $\bar{3}^+$ & $\bar{3}^-$\\
           \hline
           $\Gamma_1$  & 1  & 1                      & 1                     &  1 &  1 &  1\\
           $\Gamma_2$  & 1  & 1                      & 1                     & -1 & -1 & -1\\
           $\Gamma_3$  & 1  & $e^{\frac{2\pi i}{3}}$ & $e^{\frac{4\pi i}{3}}$ &  1 & $e^{\frac{2\pi i}{3}}$ & $e^{\frac{4\pi i}{3}}$\\
           $\Gamma_4$  & 1  & $e^{\frac{2\pi i}{3}}$ & $e^{\frac{4\pi i}{3}}$ & -1 & $e^{\frac{5\pi i}{3}}$ & $e^{\frac{\pi i}{3}}$\\
           $\Gamma_5$  & 1  & $e^{\frac{4\pi i}{3}}$ & $e^{\frac{2\pi i}{3}}$ &  1 & $e^{\frac{4\pi i}{3}}$ & $e^{\frac{2\pi i}{3}}$\\
           $\Gamma_6$  & 1  & $e^{\frac{4\pi i}{3}}$ & $e^{\frac{2\pi i}{3}}$ & -1 & $e^{\frac{\pi i}{3}}$ & $e^{\frac{5\pi i}{3}}$\\
    \end{tabular}
\end{ruledtabular}
\end{table}

\subsection{Theoretical Analyses}
\label{sec:theory}

\begin{figure}[htb]
	\begin{center}
	\includegraphics[width=0.9\columnwidth]{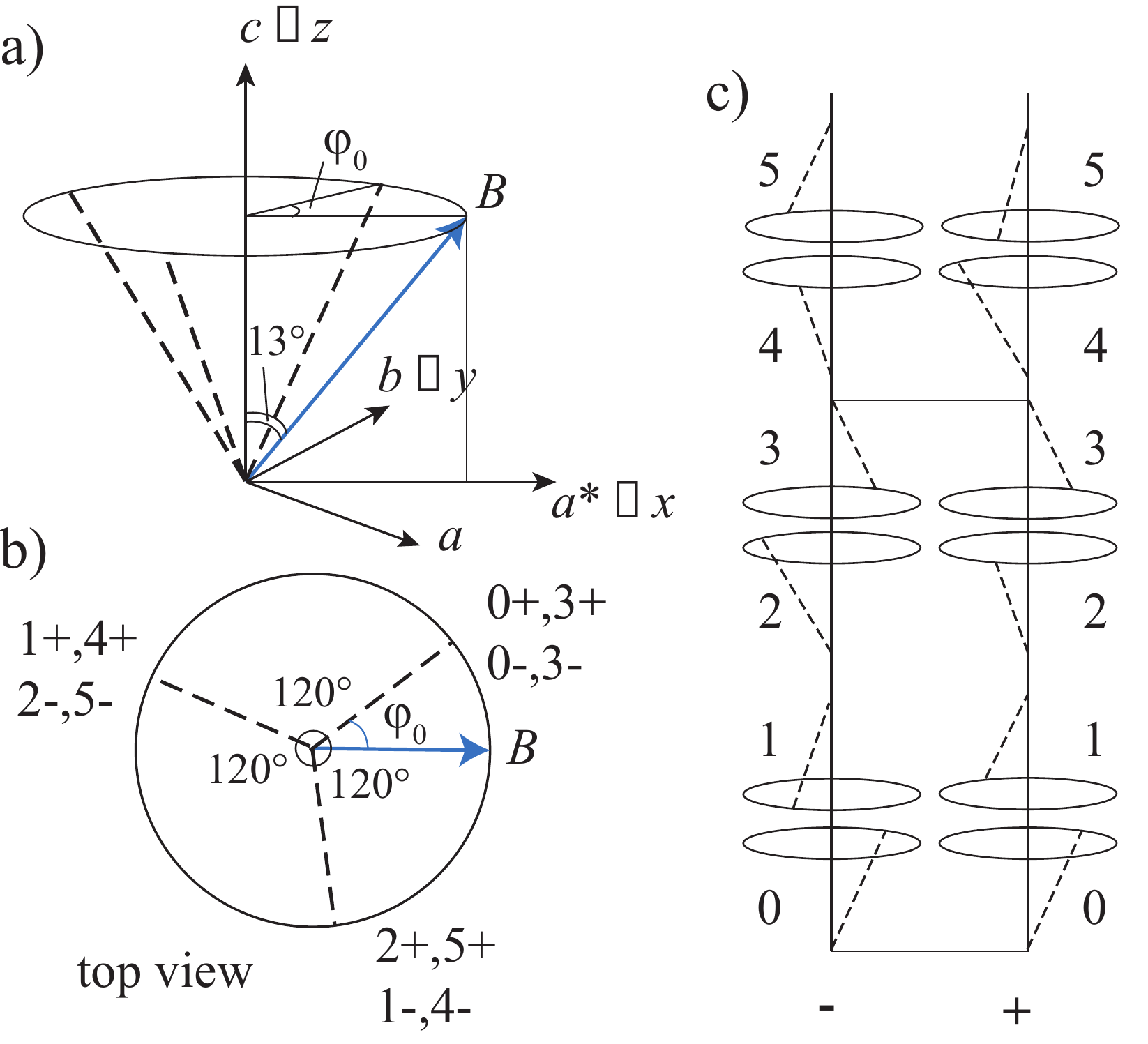}
	\caption {~a)-b) Schematic view of the magnetic field in the neutron diffraction setting (tilted by $13^\circ$ from the $c$ axis in the direction of the $a^*$ axis) and the three different local easy axes (dashed lines) in the reference frame defined by $a^*\equiv x$, $b \equiv y$, $c\equiv z$ (for convenience we display also the $a$ axis of the hexagonal coordinate system). c) Arrangement of the local easy axes along two adjacent chains of opposite helicity; here the pairs $(0,+)$-$(0,-)$ and  $(3,+)$-$(3,-)$ share a ferromagnetic bond. }
	\label{geometry}
	\end{center}
\end{figure}

As only a very limited number of magnetic reflections is accessible in the current neutron scattering experiment we turn to the theoretical modelling of the magnetic order in \GD~and its field evolution.
We modelled the magnetic interactions in \GD\ using the following Hamiltonian:
\begin{eqnarray}
\hat{H} &=& J_c\sum_{\vev{i,j}} (\hat{S}^x_i \hat{S}^x_j + \hat{S}^y_i \hat{S}^y_j + \Delta \hat{S}^z_i \hat{S}^z_j) + J_{ab} \sum_{[i,j]} \hat{\bS}_i\cdot \hat{\bS}_j \non
 && - g \sum_i \mathbf{B}_i \cdot \hat{\bS}_i
 \label{ourmodel}
\end{eqnarray}
where $\vev{\ldots}$ represents intrachain nearest-neighbor bonds and $[\ldots]$ represents interchain bonds. The interaction between spins along the chain is antiferromagnetic, of the $XXZ$ easy-axis type, $\Delta>1$, whereas the interactions in the $ab$ plane are ferromagnetic and spin-isotropic. We take into account the different orientations of the local easy axes (see Fig.~\ref{geometry}). They are tilted by $13^\circ$ with respect to the $c$-axis and their projections onto the $ab$-plane form $120^\circ$-structure. Moreover, the axis are arranged periodically along the chains (the same orientation repeats every three sites) as in Fig.~\ref{geometry}c, which displays the helicity of each chain.
The angle $\varphi_0$ is defined as as the smallest positive angle between the $a^*$-axis and  the projection of one of the easy axes.
Thus we express the external magnetic field in the coordinates of the local spin reference frame, namely $\mathbf{B}_i$. Note that the angle $\varphi_0$ in Fig.~\ref{geometry} cannot be uniquely determined by neutron diffraction; however, by performing all the relevant calculations for various values of $\varphi_0$ and orientation domains described later, we are able to see that none of the desired observables vary appreciably. Given the three different orientations of the easy axes in the material, there will be three different magnetic field vectors. 
In addition, for an antiferromagnetic material, there will be six nonequivalent spin expectation values (magnetic moments), $\mathbf{S}_\alpha$, $\alpha=0,\ldots,5$. 
More specifically, for two adjacent chains with opposite helicity, such as shown in Fig.~\ref{geometry}c, let us denote the spins at sites with the positive-helicity (negative-helicity) chain by $\hat{\bS}_{\alpha,+}$ ($\hat{\bS}_{\alpha,-}$),  $\alpha=0,\ldots,5$. 
If  the bonds $(0,+)$-$(0,-)$, $(3,+)$-$(3,-)$ are coupled by the interchain interaction, then the spin structure will be $\vev{ \hat{\bS}_{\alpha,+}} = \vev{ \hat{\bS}_{(-\alpha\mod 6),-}} = \bS_\alpha $ and will repeat periodically along the chains. 
It is worth noting that i) the sequence of magnetic moments is inverted from  one chain to the adjacent one and ii) spins coupled by ferromagnetic bonds have the same expectation value.

Having clarified symmetry constraints, we quantitatively analyzed Eq.~\ref{ourmodel} by means of Chain Mean-Field Theory (ChMFT). 
Technical details are provided in Appendix~\ref{chmftdet}. 
The first goal was to determine the parameters $J_c$, $J_{ab}$, and $\Delta$ by a least-squares optimization of the magnetization curve for $B\parallel c$ obtained by ChMFT on a $N=252$ one-dimensional cluster. In this case we can choose local spin reference frames such that $\mathbf{B}_i=B(\sin(13^\circ), 0, \cos(13^\circ) )$ $\forall i$). The estimated parameters were $J_c=6.09$ {meV} $(70.7$ {K}), $J_{ab}=-0.44\, J_c$, $\Delta=1.09$. 
This set of parameters generated an agreement within 10~$\%$ with the experimental magnetization curve as visualized in Fig.~\ref{MvsH}. 
The obtained value for $J_c$ is close to the one reported in a theoretical analysis \cite{Janson}, while the value of $J_{ab}$ lies between those reported in Refs.~\cite{Janson, Podlesnyak}. 
In Ref.~\cite{Podlesnyak} the parameters were obtained via a linear spin-wave fitting of the magnetic excitation measured by inelastic neutron scattering. 
Such a method is known to overestimate the $J_c$ by a factor of about $\frac{\pi}{2}$ (the latter being the quantum correction factor of the AF Heisenberg chain), which is close to the difference with our value. 
Also, it underestimates the anisotropy $\Delta$, which was reported to be only 1.013.

\begin{figure*}[tb]
	\begin{center}
	\includegraphics[scale=0.9]{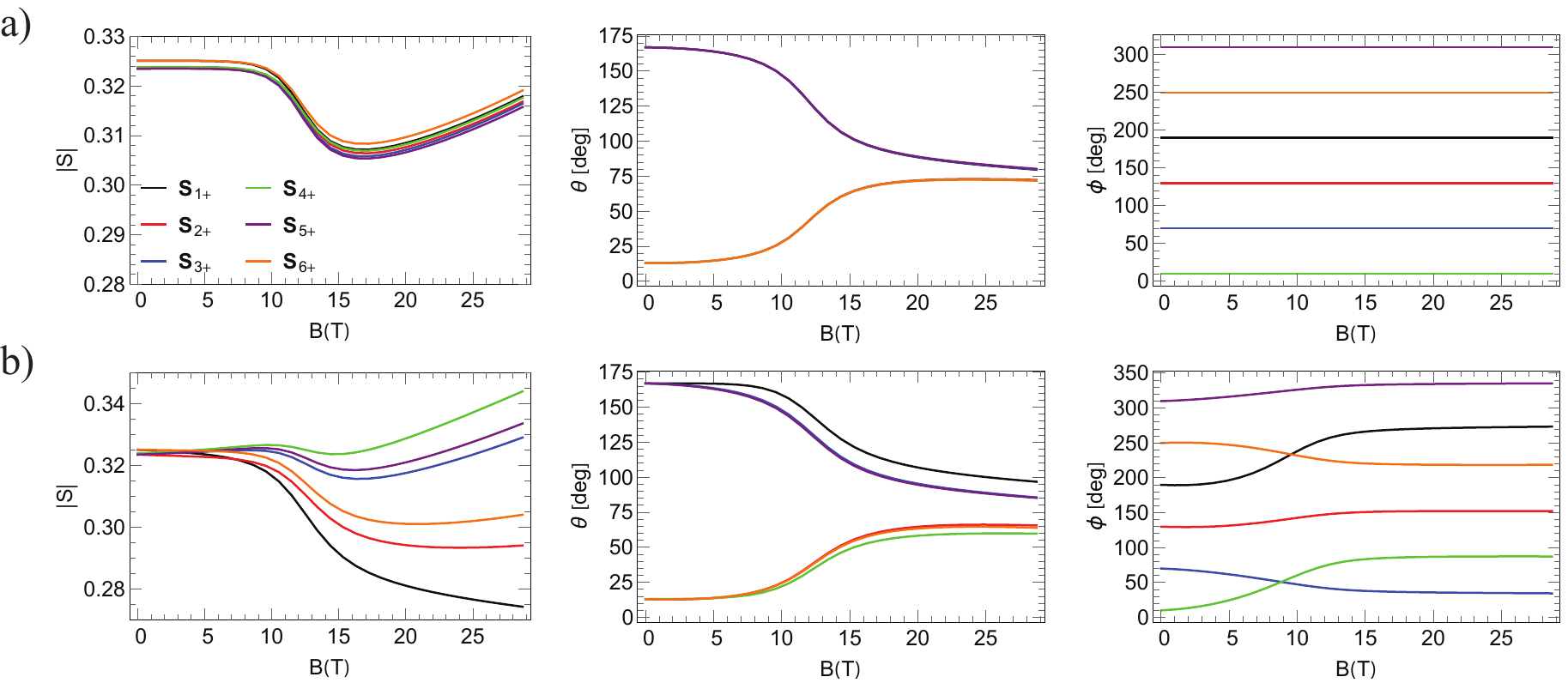}
	\caption {~Spin expectation values $\mathbf{S}_\alpha$ in spherical coordinates as a  function of the  magnetic field applied a) parallel to the $c$-axis (magnetization measurement settings) and b) 13$^\circ$ off the $c$-axis (neutron diffraction settings) ($\varphi_0=10^\circ$ is chosen). The spherical coordinate system is defined as follows: $|\mathbf{S}|$ is a spin expectation value (size of the magnetic moment), $\theta$ is an inclination of the moment with respect to the $c$-axis
	and $\phi$ is the azimuthal angle with respect to the  $a^*$-axis.}
	\label{mxmymz_B0deg}
	\end{center}
\end{figure*}

The actual microscopic spin components as a function of the magnetic field in spherical coordinates ($|\mathbf{S}|$ is the spin expectation value (the size of the magnetic moment), $\theta$ is the inclination of the moment with respect to the $c$-axis and $\phi$ is the azimuthal angle with respect to the  $a^*$-axis) are shown in Fig.~\ref{mxmymz_B0deg}a. In addition, Table~\ref{Moments} comprises numerical values for zero field and 21~T. In zero field, the system is in a N\'eel state consistent with the orientation of the easy axes. The  spin components sum up to zero pairwise and the different pairs are simply related by $120^\circ$ rotations around the $c$-axis. 
Note that the small (less than $1~$\%) discrepancies in spin magnitudes in Fig.~\ref{mxmymz_B0deg}a are merely due to the finite size (boundary effects) of the numerical calculation.
 With increasing the field the moments rotate  from the $c$-axis towards the plane as a result of a competition between the antiferromagnetic interaction and the magnetic field. As expected from the shape of $M(B)$ there is no sharp transition but about a 5~T broad range of crossover between the low and high-field states (see Fig.~\ref{mxmymz_B0deg}a).
  As can be seen from this figure, the crossover is governed by the same $\Gamma_1$ irrep of little group $G_{\bf k}$ for both the active propagation vectors, $\bf k$=(0,0,$\frac{3}{2}$) and (0,0,0). While the $S_{\perp}$ components become dominant for the former, the latter acquires a ferromagnetic $S_{\parallel}$ component. As a result, the high-field magnetic order is a 120$^\circ$-state with a finite FM-component along the $c$-axis. 
  For each field the magnetic order can be represented as field dependent linear combination of basis vectors of $\Gamma_1$ irrep with $\bf k$=(0,0,$\frac{3}{2}$) and (0,0,0). The entire magnetization process can be schematically written as  $\Gamma_1(k_z=\frac{3}{2})\rightarrow\Gamma_1(k_z=\frac{3}{2})\bigoplus\Gamma_1(k_z=0)\rightarrow\Gamma_1(k_z=0)$, where the last term is a field-induced ferromagnetic order, requiring about 100~T to be reached~\cite{Zvyagin}. 
 The high-field structures are visualized in Fig.~\ref{structure2}b-c. 

\begin{table*}
\centering
\caption{ \label{Moments} Calculated spin components (in units of $\frac{1}{2}$) along the chain at zero field and at 21~T applied parallel to the $c$-axis (magnetization measurement settings) and 13$^\circ$ off the $c$-axis (neutron diffraction settings). The reference frame shown in Fig:~\ref{geometry} is defined as $a^*\equiv x$, $b \equiv y$, $c\equiv z$.}
\begin{ruledtabular}
    \begin{tabular}{cccccccccccc}
                    & $S_x$ & $S_y$ & $S_z$ & & $S_x$ & $S_y$ & $S_z$ &  & $S_x$ & $S_y$ & $S_z$  \\
                    &  & at $B=$0~T & & & & $B=$21~T & & & & $B=21$~T & \\
                    &  & & & & & ($\parallel c$) & & & & (13$^\circ$ off the $c$-axis) & \\
           \hline
           Cu$_0$  & -0.072 &	-0.013	& -0.317 & & -0.304 &	-0.054 & 0.016 & & 0.007 & -0.270 & -0.073 \\
           Cu$_1$  & -0.047 & 0.056 & 0.315 & & -0.189 & 0.225 & 0.093 & & -0.237 & 0.124 & 0.123 \\
           Cu$_2$  & 0.025 & 0.068 & -0.316 & & 0.105 & 0.289 & 0.015 & & 0.259 & 0.185 & -0.020 \\
           Cu$_3$  & 0.072 & 0.013 & 0.316 & & 0.290 &	0.051 & 0.094 & & 0.015 & 0.283 & 0.171 \\
           Cu$_4$  & 0.047 & -0.056 & -0.315 & & 0.197& -0.235 & 0.014 & & 0.291 & -0.137 & -0.016 \\
           Cu$_5$  & -0.025 & -0.069 & 0.317 & & -0.101 & -0.278 & 0.095 & & -0.211& -0.169 & 0.132 \\
           
    \end{tabular}
\end{ruledtabular}
\end{table*}

Two caveats must be considered with respect to the above analysis. 
First, ChMFT was performed at $T=0$. This is not a significant problem as the temperature of the magnetization measurement is only $2\%$ of the exchange coupling $J_c$. 
Second, we utilized the saturation field value ($H_s$) of $\sim 100$~T \cite{Zvyagin}  and related two of our parameters via the approximation $g \mu_B H_s \simeq J_c(1+\Delta)$ (note that the equality would be exact if the local easy axes were directed along the $c$-axis, so that the magnetic field contained only a longitudinal component).

Given the above optimized parameters, it is now possible to perform the ChMFT calculation for the neutron experiment, (Fig.~\ref{scat_geom}), in which the magnetic field lies in the $a^*$-$c$ plane and forms an angle $\omega_s=13^\circ$ with the $c$-axis. 
The values of the magnetic moments $\bS_\alpha$, obtained in the reference frame defined above, are displayed in Fig.~\ref{mxmymz_B0deg}b. 
As the magnetic field is progressively increased, the higher symmetry of the N\'eel state is lost due to the orientation of the field and all the six nonequivalent $\bS_\alpha$ values become apparent. We may call this a ``deformed N\'eel state''. Around $B^*=12.2$ T the spin structure crosses over smoothly to a high-field ``deformed spin-flop state''. 
Both structures are visualized in Fig.~\ref{structure2}d-e. 
The magnetic moments undergo the biggest quantum depletion (reduction in the moment size $|\bS_\alpha|$ from its classical value 1/2) in the vicinity of $B^*$, more specifically at the onset of the deformed spin-flop state (see Fig.~\ref{mxmymz_B0deg}, left panels, where the $\omega_s=0^\circ$ ($B\parallel c$) and $\omega_s=13^\circ$ settings are compared). 
This is due to the fact that quantum fluctuations are enhanced by the competition between the spin-anisotropy and the magnetic field, whose magnitudes are comparable in this region.
Moreover, drawing from the analogy with N\'eel and spin-flop states, we can argue that the deformed spin-flop state has softer excitations and therefore is more subject to  reduction of the moments.

Coming to the symmetry of the field-induced state for the field applied off the $c$-axis, one should follow a similar procedure as described above for the $B\parallel c$ case. However, here it will not be sufficient to combine only the $\Gamma_1$ irreps of the active propagation vectors. Indeed, as can be seen from Fig.~\ref{mxmymz_B0deg}b, one would require six nonequivalent sites (moment sizes) to describe the high-field structure. One should consider also other inversion-even irreps, namely $\Gamma_3$ and $\Gamma_5$, for both propagation vectors.

Even though it is possible, in principle, to use ChMFT at finite temperature, severe limitations on the accessible system size arise (they are reminiscent of those in finite-temperature exact diagonalization). At the same time, the model shown in Eq.~\ref{ourmodel} is not amenable to Quantum Monte Carlo (QMC) simulations, because the $XXZ$ anisotropy together with the transverse component of the magnetic field (unavoidable at 2/3 of the sites, at least) brings about the well-known sign problem~\cite{Syljuasen, Melko}. Therefore, we chose a simplified model in which all easy-axes are aligned with the $c$-axis and the magnetic field has only longitudinal component, which in practice is equivalent to setting $\mathbf{B}_i=(0,0,B)$ $\forall i$ in Eq.~\ref{ourmodel}. We  performed QMC simulations using the DSQSS package \cite{DSQSS} on a system of size $N=5184$, namely $12^2$ coupled chains of length 48. 
In this system, the most convincing estimate of the couplings is in the case $J_c=5.98$ {meV} ($69.4$ {K}), $J_{ab}=-0.4\, J_c$, $\Delta=1.13$. 
In Fig.~\ref{MvsT} we report the comparison of this model with experimental measurement of the curve $M(T)/B$ at $B=7$ T, $B\parallel c$. 
The agreement within 4$\%$ is quite reasonable, given the oversimplification of the model, although a  better agreement can be obtained by choosing the spin-isotropic model ($\Delta=1$) with $J_{ab}=-0.5 J_c$ \cite{Janson}. 
However, the latter model cannot be justified since it does not present any transition or visible crossover before saturation in the magnetization curve $M(B)$. Moreover, it does not account for the experimentally observed gap in the excitation spectrum~\cite{Podlesnyak}.

\section{Discussion}
\label{discussion}

For the zero-field state, our theoretical model (Eq.~\ref{ourmodel}), consisting of AFM spin chains with relatively small interchain FM coupling,  clearly reproduces the experimental results of Belokoneva~et~al.~\cite{Belokoneva}. 
The Cu$^{2+}$ spins in the chains, which are  reduced by quantum fluctuations by about one third, are  N\'eel-ordered (Fig.~\ref{structure2}a). 
However, while their projection along the $c$-axis is predominant, the chains do have in-plane components due to the tilting of the magnetic easy axes.

Application of an external magnetic field produces a smooth deformation of  the spin structure, in which the six sublattice magnetizations gradually rotate into a deformed N\'eel state and subsequently, through a smeared spin-flop transition, into a deformed spin-flop state at higher fields (Fig.~\ref{structure2}b-c). 
While this might be somewhat counterintuitive for a system of non-frustrated coupled AFM chains with easy-axis spin anisotropy, in which a sharp spin-flop transition might be envisioned, it is confirmed by all the experimental observations, in particular by the magnetization curve for $B\parallel c$ (Fig.~\ref{MvsH}) and the intensity of the (-4,1,$\frac{1}{2}$) magnetic reflection as a function of the magnetic field (Fig.~\ref{mag_ref}) where $B$ is tilted off the $c$-axis by $\omega_s = 13^\circ$. 
In the latter case, using the theoretically derived spin configurations, described in the previous section, the magnetic structure factors for the covered magnetic field range (0 - 21~T) have been simulated using FullProf~\cite{FullProf}. The calculated magnetic intensities are plotted in Fig.~\ref{mag_ref}a in comparison with the measured ones. 
The only refined parameter in the calculated curve is a scaling factor while the intensities have been obtained from the spin configurations (Fig.~\ref{mxmymz_B0deg}).
For the intensity calculations we assume that the magnetic structures over the entire field range are represented by three equally populated orientation domains. 
Their magnetic moments have equivalent directions with respect to $\bf{k_M}$ and according to the hexagonal symmetry. 
These oriented domains contribute to the same magnetic reflections. 
The agreement between the experiment and the calculations is better than 10~$\%$, providing a microscopic basis of the scenario described.  

\begin{figure}[bt]
	\begin{center} 
		\includegraphics[width=0.95\columnwidth]{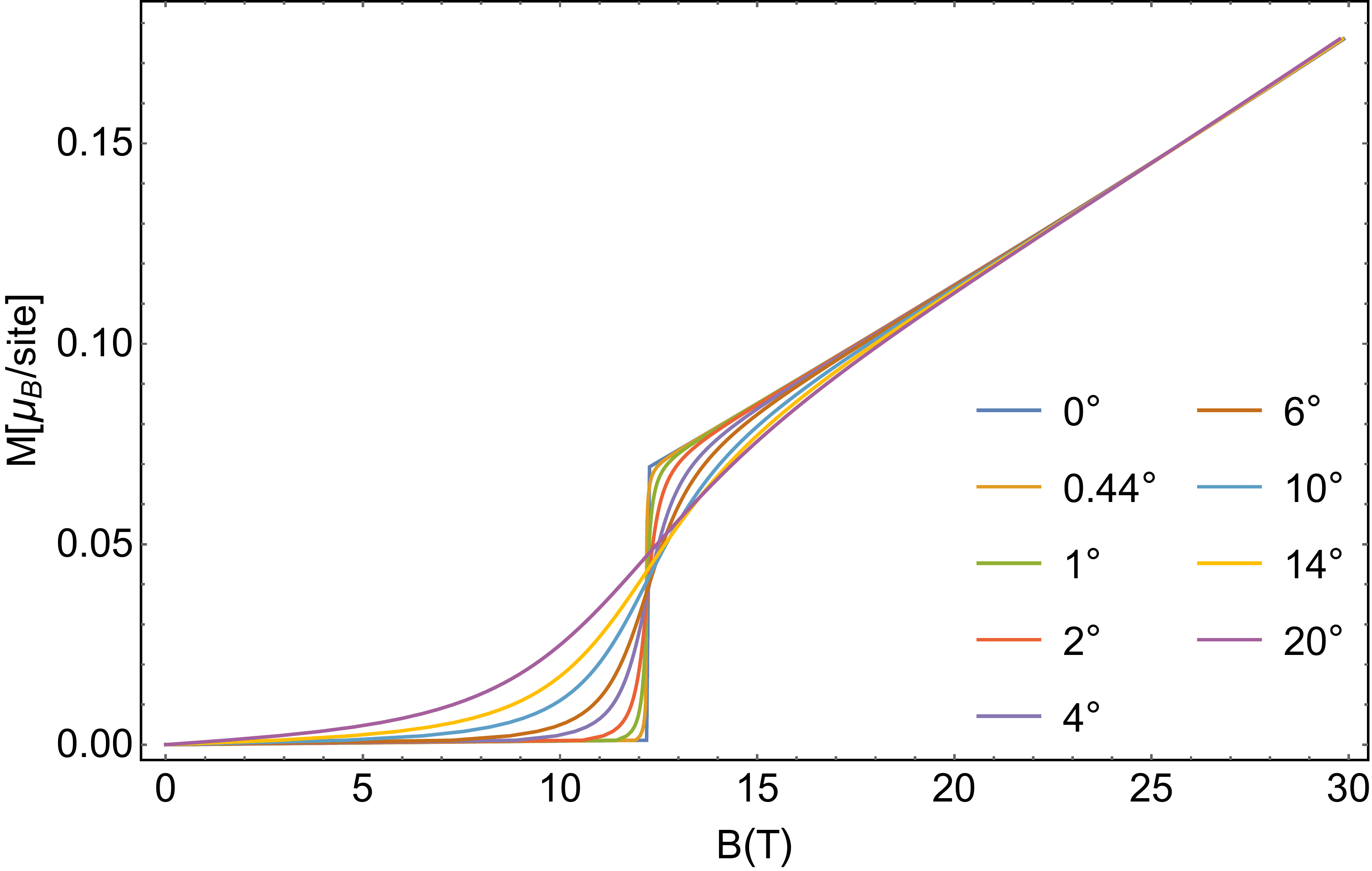} 
		\caption{~Magnetization curve $M(B)$ of the model in Eq.~\ref{toyxxz} for
		$J_{ab}=-0.44\, J_c$, $\Delta=1.09$ and various values of $\theta$. The maximum angle for which a first-order transition takes place is $\theta_c=0.44^\circ$.
		}
		\label{degdep}
	\end{center}
\end{figure}

The main question arising at this point is whether our intentional field offset with respect to the $c$-axis is responsible for the observed crossover. 
In other words, whether a "classical" spin-flop transition can be realized in \GD\ at any orientation of the magnetic field.
To understand this in a simpler and physically transparent way, we evaluated the effect of a transverse component of the magnetic field in a standard $XXZ$ model (the easy axes at all sites are all aligned). The Hamiltonian we considered is
\begin{eqnarray}
\hat{H}_{XXZ} &=& J_c\sum_{\vev{i,j}} (\hat{S}^x_i \hat{S}^x_j + \hat{S}^y_i \hat{S}^y_j + \Delta \hat{S}^z_i \hat{S}^z_j) + J_{ab} \sum_{[i,j]} \hat{\bS}_i\cdot \hat{\bS}_j \non
 && - B \cos(\theta) \sum_i  \hat{S}^z_i - B \sin(\theta) \sum_i  \hat{S}^x_i,
 \label{toyxxz}
\end{eqnarray}
where $B$ is the magnetic field amplitude and $\theta$ is the angle between the easy-axis and the magnetic field direction. 
We fixed $\Delta=1.09$, $J_{ab}=-0.44\, J_c$ as in green dioptase and calculated the magnetization curve $M(B)$ as a function of $\theta$.  
The results of these calculations are summarized in Fig.~\ref{degdep}. As can be seen, the first-order spin-flop transition exists only for $0<\theta<\theta_c$, with $\theta_c\simeq 0.44^\circ$, whereas for $\theta>\theta_c$ there is only a crossover between the low-field and the high-field states. 
In other words, the $B$-$\theta$ phase diagram contains a first-order line ending at a critical point, similar to a liquid-gas transition.

In green dioptase, as better described by Eq.~\ref{ourmodel}, the intrinsic misalignment of the local easy axes, which have three nonequivalent directions depending on the site, is such that the transverse component of the magnetic field is always too large, at least at 2/3 of the sites, for a sharp first-order transition to take place. Thus, we suggest that a crossover between low- and high-field states must occur for any orientation of the external magnetic field. Indeed, in Fig.~\ref{mag_ref}a we plot the calculated intensity of the (-4,1,$\frac{1}{2}$) reflection for the high symmetry direction $B\parallel c$. As expected, even in this case it shows a smooth variation of magnetic intensity as function of field contrary to a classical first-order spin flop transition.

\begin{figure}[bt]
	\begin{center} 
		\includegraphics[width=0.9\columnwidth]{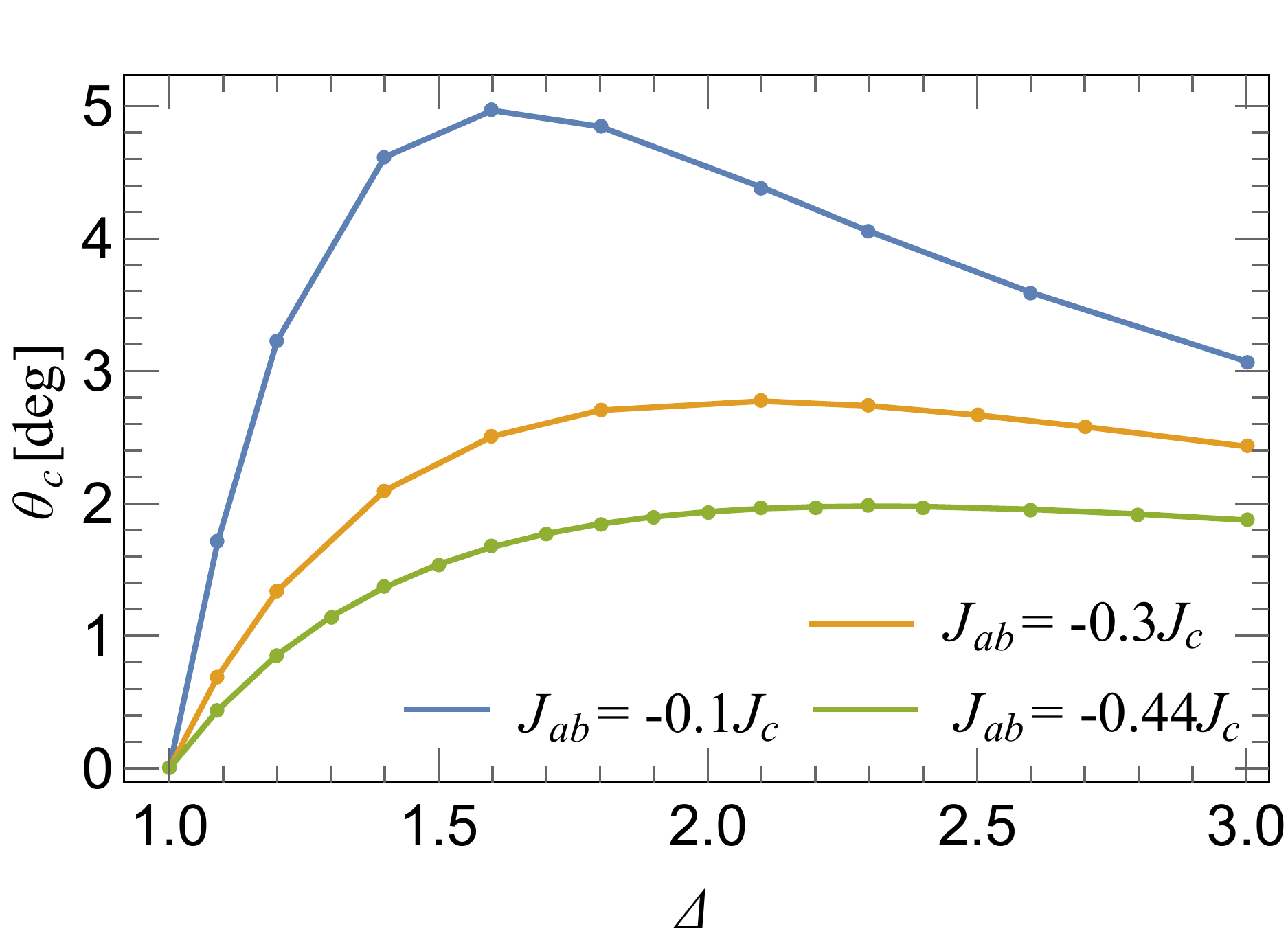}
		\caption{~Dependence of the critical angle $\theta_c$ on the spin anisotropy $\Delta$ at various values of the interchain coupling
		$J_{ab}$ for the model Eq.~\ref{toyxxz}. The lines are just guides for the eye.
		}
		\label{thetac}
	\end{center}
\end{figure}

In order to determine dependence of $\theta_c$ on the spin anisotropy $\Delta$ and interchain coupling $J_{ab}$ we again employed the simple model given in Eq.~\ref{toyxxz}. We use it within the domain of validity of ChMFT, namely $J_{ab}$ is of order $-0.1 J_c$.  Note that when $J_{ab}$ is much smaller than that, strong quantum fluctuations may stabilize incommensurate orders \cite{Okunishi2007}. 
The results of the calculation are summarized in Fig.~\ref{thetac}. 
Two salient features are apparent. 
Firstly, $\theta_c$ has a maximum at intermediate $\Delta$. 
This is expected, because no sharp spin-flop transition exists in the isotropic case ($\Delta=1$) and in the  Ising limit ($\Delta \to \infty$). 
Secondly, a smaller interchain coupling appears to make the sharp spin-flop transition slightly more stable against a transverse component of the magnetic field. 
The latter aspect can be understood if one considers that a non-vanishing $\theta_c$ is a purely quantum effect (classically $\theta_c=0$ in the whole parameter space \cite{Rohrer1969}). When $J_{ab}$ weakens  (the dimensionality of the system is effectively reduced)  quantum fluctuations become more important and the value of $\theta_c$ increases.

\section{Conclusions}

Magnetic order and magnetic properties of \GD~in high magnetic field have been studied by means of neutron diffraction and magnetization measurements in magnetic fields up to 30~T applied close to or along the $c$-axis. Both experimental techniques show a smooth crossover around $B^*=12.2$~T at 1.5~K. The results can be explained in terms of a modified  model of quantum $XXZ$ spin chains coupled by ferromagnetic bonds in the transverse plane. 
The crucial components of this model are the different orientations of the local easy axes (the same orientation repeating every three sites along the chain) and the  opposite helicities of adjacent chains. As a consequence of this, the spins (or the majority thereof) always feel a sizeable component of the magnetic field transverse to their easy axis.
Thus, even when the applied magnetic field is (nearly) parallel to the $c$-axis, no sharp transition takes place in going from the low-field (deformed N\'eel) to the high-field (deformed spin-flop) state, but a strong crossover appears around $B^*=12.2$~T. 
In order to understand this phenomena more generally, we developed a more traditional model of coupled $XXZ$ easy-axis quantum spin chains with aligned easy axes and looked for the maximum angle $\theta_c$ at which the magnetic field can be tilted to observe a first-order spin-flop transition. 
This critical angle, which vanishes at the classical level in our model, is greatest at intermediate spin anisotropies (between the isotropic and the Ising limits) and is enhanced by quantum fluctuations when the interchain coupling becomes smaller. 
These considerations provide useful directions for future experiments with analogous spin-anisotropic materials. 

\section*{Acknowledgments}
We greatly acknowledge R. Wahle, S. Gerischer, S. Kempfer, P. Heller and P. Smeibidl for their support at the HFM/EXED facility at the Helmholtz-Zentrum Berlin. O.P. acknowledges support by ICC-IMR, Tohoku University. S.E.N. acknowledges support from the International Max Planck Research School for Chemistry and Physics of Quantum Materials (IMPRS-CPQM). D.Y. was supported by KAKENHI from Japan Society for the Promotion of Science, Grant Number 18K03525 (D.Y.), and ``Early Eagle'' grant program from Aoyama Gakuin University Research Institute. G.M. thanks Y.~Motoyama for useful correspondence. This research used resources at the Spallation Neutron Source, a DOE Office of Science User Facility operated by the Oak Ridge National Laboratory. This material is based upon work supported by the U.S. Department of Energy, Office of Science, Office of Basic Energy Sciences, Chemical Sciences, Geosciences, and Biosciences Division. Powder x-ray diffraction measurements were conducted at the Center for Nanophase Materials Sciences (CNMS) (CNMS2019-R18) at the Oak Ridge National Laboratory (ORNL), which is a DOE Office of Science User Facility. This work was partly supported by the Deutsche Forschungsgemeinschaft, through ZV 6/2-2, as well as by the HLD at HZDR, member of the European Magnetic Field Laboratory (EMFL).

\appendix
\section{Details of Chain Mean-Field Theory (ChMFT)}
\label{chmftdet}

The idea of ChMFT, and cluster mean-field theories in general, is to reduce a many-body problem, such as Eq.~\ref{ourmodel}, to a problem on a finite-size cluster supplemented by mean-field boundary conditions. 
Those mean fields appear in the form of effective magnetic fields to which the spins on the edge of the cluster are subject, and have the purpose of including the effects of  correlation with spins outside the cluster. 
The solution is then obtained by an iterative procedure in which the cluster is treated in  full quantum-mechanical fashion [by exact diagonalization or density matrix renormalization group (DMRG)] and the mean fields are made to satisfy self-consistency equations \cite{Yamamoto_PRB86}. 
This method has proven quite successful, in particular in the treatment of frustrated antiferromagnets, for which the QMC method is not applicable due to sign problem \cite{Yamamoto_PRL112,*Yamamoto_PRL112e,Yamamoto_PRL114,Yamamoto_2016,Yamamoto_2017,Yamamoto_2019}.

After the symmetry considerations in Sec.~\ref{sec:theory}, in the case of  Eq.~\ref{ourmodel} we can choose the finite cluster as a one-dimensional segment of length $N$, which must be a multiple of 6. The ChMFT Hamiltonian reads
\begin{eqnarray}
\hat{H}_C &=& J_c\sum_{i=0}^{N-2} (\hat{S}^x_i \hat{S}^x_{i+1} + \hat{S}^y_i \hat{S}^y_{i+1} + \Delta \hat{S}^z_i \hat{S}^z_{i+1}) 
\non
 && - g \sum_{i=0}^{N-2} \mathbf{B}_i \cdot \hat{\bS}_i -  \sum_{i=0}^{N-2} \mathbf{b}_i^{MF} \cdot \hat{\bS}_i,
 \label{ourmodelcmf}
\end{eqnarray}
where $\mathbf{b}_i^{MF}$ are the mean-fields, coming from the mean-field decoupling $\hat{S}^A_i \hat{S}^A_j = \hat{S}^A_i \vev{S}^A_j +\vev{S}^A_i\hat{S}^A_j-\vev{S}^A_i\vev{S}^A_j$, $A=x,y,z$. Specifically, they will be given by
\begin{eqnarray}
&\mathbf{b}_i^{MF} = J_{ab} \mathbf{S}_{(i\mod 6)} \\
&\mathbf{b}_0^{MF}  = J_{ab} \mathbf{S}_0 + J_c (S^x_5,S^y_5,\Delta S^z_5) \\
&\mathbf{b}_N^{MF} = J_{ab} \mathbf{S}_5 + J_c (S^x_0,S^y_0,\Delta S^z_0),
\end{eqnarray}
with the expectation values $\mathbf{S}_\alpha$, $\alpha=0,\ldots,5$, defined as in Sec.~\ref{sec:theory}. Although a relative variation of approximately 5~$\%$ between the $g_{\parallel}$- and $g_{\perp}$-factors has been reported in the literature~\cite{Ohta2009}, the tilting of the magnetic field in our experiments does not exceed 13$^\circ$. Therefore we neglect possible variations of $g$ and fix it to 2.2 throughout all the calculations.
In the specific setting used in this paper, we set $N=252$ and solve the cluster problem via DMRG calculation performed with the ITensor package \cite{ITensor}. The magnetic moments are then recalculated as
\begin{equation}
    \mathbf{S}_\alpha = \frac{6}{N} \sum_{l=0}^{(N/6)-1} \frac{\Tr{(\hat{\mathbf{S}}_{\alpha+6l} \, e^{\beta \hat{H}_C})}}{\Tr{( e^{\beta \hat{H}_C})}}, \quad \alpha=0,\ldots,5,
    \label{average}
\end{equation}
and substituted back into $\mathbf{b}_i^{MF}$. For us, it sufficed to take the zero temperature ($\beta=\infty$) limit, as argued in Sec.~\ref{sec:theory}.
In this case, only the ground state contributes to the trace in \ref{average}. 
The procedure is repeated until convergence.

Below we describe how to calculate the magnetic field $\mathbf{B}_i$ in the local reference frame.  Our choice of the ``laboratory frame'' is $a^*\equiv x$, $b \equiv y$, $c\equiv z$. In this frame the external magnetic field in the setup of Fig.~\ref{scat_geom} is $\mathbf{B}=(\sin (13^\circ),0,\cos (13^\circ))$. The local spin reference frame is defined as follows. The local $z$ direction  coincides with the local easy axis. Referring to Fig.~\ref{geometry} we  have
\begin{equation}
    \hat{z}_{i,\pm} =
\begin{pmatrix}
\sin(13^\circ) \cos\phi_{i,\pm}\\
\sin(13^\circ) \sin\phi_{i,\pm} \\
\cos(13^\circ)
\end{pmatrix} \qquad i=0,\ldots,5
\end{equation}
and $\phi_{i,\pm}=\varphi_0\pm  120^\circ \times i$. 
As mentioned in Sec.~\ref{ourmodel}, $\varphi_0$ cannot be uniquely determined by any available experimental data, so our strategy was to perform the calculations for several values between $0^\circ$ and $60^\circ$ (symmetry arguments account for values outside this range). 
None of the calculated observables, most importantly the intensity of the magnetic reflections averaged over three orientation domains, showed any important difference as $\varphi_0$ was changed. 
The angle between the magnetic field and the local easy axis is determined by
\begin{equation}
    \cos\alpha_{i,\pm} = \hat{z}_{i,\pm} \cdot \mathbf{B} /B = (\cos(13^\circ))^2 + (\sin (13^\circ))^2  \cos(\phi_{i,\pm})  
\end{equation}
We then choose the local $\hat{x}_{i,\pm}$ and $\hat{y}_{i,\pm}$ direction in such a way that in the local frame
\begin{equation}
\mathbf{B}_{i,\pm}  = B \begin{pmatrix}
 \sin\alpha_{i,\pm} \\
0\\
 \cos\alpha_{i,\pm}
\end{pmatrix}.
\end{equation}
In practice there are three different local magnetic field vectors (dictated by the three directions of the easy axes), whose sequencing along the chain can be read off Fig.~\ref{geometry}.


%

\end{document}